\documentclass[aps,prd,preprintnumbers,amsmath,amssymb,nofootinbib,11pt]{revtex4}
\usepackage{eepic}
\usepackage{indentfirst}
\usepackage{mathrsfs}
\usepackage{fancyhdr}
\usepackage{ulem}
\usepackage{float}
\usepackage{graphicx}
\usepackage[
colorlinks=true,
linkcolor=black,
breaklinks=true,
urlcolor=blue,
citecolor=green]{hyperref}
\usepackage{epstopdf}
\usepackage{bm,bbm}

\usepackage{xcolor}

\usepackage{tabularx}
\usepackage{subfigure}
\usepackage{epsfig}
\usepackage{color}
\usepackage{slashed}
\usepackage{hyperref}

\newcommand{\be}{\begin{equation}}
\newcommand{\ee}{\end{equation}}
\newcommand{\ba}{\begin{array}{c}}
\newcommand{\ea}{\end{array}}
\renewcommand{\L}{\mathscr{L}}
\newcommand{\M}{\mathscr{M}}

\newcommand{\bra}{\langle}
\newcommand{\ket}{\rangle}
\newcommand{\nn}{\nonumber}
\newcommand{\MeV}{\,\text{MeV}}
\newcommand{\GeV}{\,\text{GeV}}

\begin{document}

\title{\boldmath Chromopolarizability of charm-beauty quarkonium from $B_c(2S) \rightarrow B_c\pi\pi$ transition   }

\author{ Yun-Hua~Chen}\email{yhchen@ustb.edu.cn}
\author{ Zi-Wei Li}
\affiliation{School of Mathematics and Physics, University
of Science and Technology Beijing, Beijing 100083, China}

\begin{abstract}

The chromopolarizability of a charm-beauty quarkonium characterizes its interaction with soft gluonic fields
and can be probed through heavy quarkonium decays. Using the dispersion theory which considers the $\pi\pi$ final state interaction model-independently, we analyze the transition
$B_c(2S) \rightarrow B_c\pi^+\pi^-$ and determine the chromopolarizability $\alpha_{B_c(2S) B_c}=(0.66\pm 0.25)$ GeV$^{-3}$ and the parameter $\kappa =-0.02\pm 0.13$.
Our results indicate that the transitional chromopolarizability of the $c\bar{b}$ state lies between the transitional chromopolarizabilities of the $c\bar{c}$ and $b\bar{b}$ states.
These findings provide valuable insights into the interactions of charm-beauty quarkonium with light hadrons.

\end{abstract}


\maketitle

\newpage

\section{Introduction}

The study of the $B_c$ system is important since it is the only one that contains two different heavy flavors in the Standard Model. The $c\bar{b}$ states, unlike their counterparts in the $J/\psi$ and $\Upsilon$ families, cannot annihilate into gluons, making them more stable.
The ground $B_c(1^1S_0)$ state was discovered by the CDF Collaboration at the Tevatron collider in 1998~\cite{CDF:1998ihx,CDF:1998axz}. In 2014, the ATLAS Collaboration observed a peak at $6842\pm4\pm5 $ MeV~\cite{ATLAS:2014lga},
which could be interpreted either as the $B_{c}^{*}(2^3S_{1})$ excited state or a pair of peaks resulting from the decays of $B_{c}(2^{1}S_{0}) \rightarrow B_{c}(1^{1}S_{0})\pi^{+}\pi^{-} $
and $B_{c}^{*}(2^{3}S_{1}) \rightarrow B_{c}^{*}(1^{3}S_{1})\pi^{+}\pi^{-} $ followed by  $B_{c}^{*}(1^{3}S_{1}) \rightarrow B_{c}(1^{1}S_{0})\gamma $. In 2019, the CMS~\cite{CMS:2019uhm} and LHCb~\cite{LHCb:2019bem} Collaborations identified signals consistent with the $B_{c}(2S)$ and $B_{c}^{*}(2S)$ states in the $B_{c}(1S)\pi^{+}\pi^{-}$ invariant mass spectrum. Shortly thereafter, the normalized invariant mass distributions of the pions pair emitted in the $B_{c}^{(*)}(2S) \rightarrow B_{c}^{(*)}\pi^{+}\pi^{-} $ decays were presented~\cite{CMS:2020rcj}. The dipion transition to the ground $B_c$ state is the dominant decay mode of the $B_{c}(2S)$ state and will be key to characterizing those states, which we focus on in this study.
The dipion transitions between heavy quarkonia are typically described as a two-step process in which the soft gluons are first emitted from the heavy quarks and then recombine into light quarks.
The heavy quarkonia are expected to be compact, and the
method of QCD multipole expansion~\cite{Voloshin:1980zf,Novikov:1980fa,Kuang:1981se,Kuang:2006me}, combined with soft-pion theorems, has been successfully used to study dipion transitions between the equal-mass $Q\bar{Q}$ systems, i.e., $\psi(2S) \rightarrow \psi(1S)\pi\pi $ and $\Upsilon(2S) \rightarrow \Upsilon(1S)\pi\pi$. Since the electricmultipole transitions are only sensitive to the relative position of the quark and antiquark, the same method has been used to study the hadronic transition for the unequal-mass $Q\bar{Q}^\prime$ systems $B_c(2S) \rightarrow B_c(1S)\pi\pi$~\cite{Eichten:1994gt,Godfrey:2004ya,Eichten:2019gig,Martin-Gonzalez:2022qwd,Li:2023wgq,Gao:2024yvz}. However, the $\pi\pi$ final-state
interaction (FSI) was not considered in the Refs.~\cite{Eichten:1994gt,Godfrey:2004ya,Eichten:2019gig,Martin-Gonzalez:2022qwd,Li:2023wgq,Gao:2024yvz}. In the $B_c(2S) \rightarrow B_c(1S)\pi\pi$ process, the dipion invariant mass reaches nearly 600~MeV, making the $\pi\pi$ FSI significant and requiring careful consideration. We will employ dispersion theory to account for the $\pi\pi$ FSI and study the decay $B_c(2S) \rightarrow B_c(1S)\pi\pi$. The same method has been successfully used to describe transitions of $\psi(2S) \rightarrow \psi(1S)\pi\pi $ and $\Upsilon(2S) \rightarrow \Upsilon(1S)\pi\pi$~\cite{Chen:2015jgl,Chen:2016mjn,Chen:2019gty,Chen:2019hmz}.

The chromopolarizability $\alpha$ of a heavy quarkonium quantifies its effective interaction with soft gluons and is an important parameter for describing quarkonium interactions. Recently, the chromopolarizability of heavy quarkonium has garnered increased interest due to its relevance in interpreting the structures of multiquark hadrons containing heavy quark and antiquark.
For reasonable values of the chromopolarizabilities of the charmonia, bound states such as the double-$J/\psi$ state, several hadro-charmonium bound states, and baryo-charmonium bound states have been identified with certain $XYZ$ states and the $P_c^+$ pentaquark states~\cite{Dong:2021lkh,Voloshin:2007dx,Dubynskiy:2008mq,Sibirtsev:2005ex,Eides:2015dtr,Tsushima:2011kh}.
Similarly, for appropriate values of the chromopolarizabilities of bottomonia, several hadro-bottomonium and baryo-bottomonium bound states have been predicted~\cite{Ferretti:2018kzy,Ferretti:2018ojb}. The $c\bar{b}$ states are intermediate between the $c\bar{c}$ and $b\bar{b}$ states, both in mass and in size. Therefore, it is expected that the values of the chromopolarizabilities of the $c\bar{b}$ states will fall between the values of the chromopolarizabilities of the $c\bar{c}$ and the $b\bar{b}$ states. For certain values of the chromopolarizabilities of the $c\bar{b}$ states, multiquark hadrons containing a $c\bar{b}$ component may also exist.
While the diagonal chromopolarizabilities $\alpha_{AA}$, with $A$ representing a heavy quarkonium, cannot be determined directly from current experimental data. One possible method to estimate $\alpha_{AA}$ assumes that the heavy quarkonia are purely color-Coulombic systems. This assumption works well for ground-state bottomonia but introduces significant uncertainties for charmonia and excited bottomonia~\cite{Ferretti:2018ojb,Dong:2022rwr}.
On the other hand, the determination of the transitional chromopolarizability $\alpha_{A^\prime A}\equiv\alpha_{A^\prime\to A}$ is of importance as it serves as a reference benchmark for each of the diagonal terms due to the Schwartz inequality: $\alpha_{AA}\alpha_{A^\prime A^\prime}\geq \alpha_{A^\prime A}^2$~\cite{Voloshin:2004un,Sibirtsev:2005ex}.
In this work, we will extract the value of $\alpha_{B_c(2S)B_c}$ from the $B_c(2S) \rightarrow B_c(1S)\pi\pi$ transition.

The theoretical framework is outlined in Sec.~\ref{theor}. In Sec.~\ref{pheno}, we fit the decay
amplitudes to the data of the $\pi\pi$ invariant mass distribution of the $B_c(2S) \rightarrow B_c(1S)\pi\pi$ process, and determine the chromopolarizability $\alpha_{B_c(2S)B_c}$ and the parameter $\kappa$. A summary is provided in Sec.~\ref{conclu}.

\section{Theoretical framework}
\label{theor}

The low energy interaction of heavy quarkonia with light hadrons
is mediated by soft gluons. The heavy quarkonia are expected to be compact, so that the method of QCD
multipole expansion is often employed to study the interaction of heavy quarkonia with
soft gluons. The leading $E1$ chromo-electric dipole
term in the expansion is simple
$H_{E1}=-\frac{1}{2}\xi^a \textit{\textbf{r}}\cdot \textit{\textbf{E}}^a$,
where $\xi^a=t_1^a-t_2^a$ is the difference between the SU(3) color generators acting
on the quark and anti-quark, $\textit{\textbf{r}}$ is the relative position vector between the quark and anti-quark, and $\textit{\textbf{E}}^a$ is
the chromoelectric field.
For the transition between two $S$-wave heavy quarkonium
states $A$ and $B$ with an emitted gluon system which possesses the
quantum numbers of two pions, the leading order amplitude is in the second order of $H_{E1}$~\cite{Voloshin:2007dx}
\begin{equation}
\langle B|H_{eff}|A\rangle=\frac{1}{2}\alpha_{AB} \textit{\textbf{E}}^a\cdot
\textit{\textbf{E}}^a,,
\end{equation}
where $\alpha_{AB}$ denotes the chromopolarizability, quantifying the strength of the $E1$-$E1$ chromo-electric
interaction of the quarkonium with the soft gluon system.

The amplitude for the dipion transition between two $S$-wave heavy quarkonium states, $A$ and $B$,
is expressed as~\cite{Voloshin:2007dx,Brambilla:2015rqa}
\begin{equation}\label{eq.MultipoleAmplitude}
M_{AB}=2\sqrt{m_Am_B}\alpha_{AB}\langle
\pi^+(p_c)\pi^-(p_d)|\frac{1}{2}\textit{\textbf{E}}^a\cdot \textit{\textbf{E}}^a|0\rangle=\frac{8\pi^2}{b}\sqrt{m_A m_B}\alpha_{AB}(\kappa_1 p^0_c p^0_d-\kappa_2 p^i_c p^i_d),
\end{equation}
where the factor $2\sqrt{m_Am_B}$ arises from the relativistic
normalization of the amplitude. $b$ denotes the first coefficient of the QCD beta function,
and its expression is $ b=\frac{11}{3}N_c-\frac{2}{3}N_f,$ where $N_c=3$ and $N_f=3$ are the number of colors and of light flavors, respectively. The parameters $\kappa_1$
and $\kappa_2$ are not independent, since $\kappa_1=2-9\kappa/2$ and $\kappa_2=2+3\kappa/2$, where the value of $\kappa$ can be determined from experimental data.

The above expression of the QCD multipole expansion together with the
soft-pion theorem, can be derived by constructing a chiral
effective Lagrangian for the $B_c(2S) \rightarrow B_c(1S)\pi\pi$
transition~\cite{Mannel,Chen:2016mjn}
\begin{equation}\label{LagrangianPsiPrimePsipipi}
\L_{B_c(2S)B_c\pi\pi}
= c_1 B_c(2S)B_c  \bra u_\mu u^\mu\ket
+c_2 B_c(2S)B_c\bra J^\dagger J^\prime \ket \bra u_\mu u_\nu\ket v^\mu v^\nu
+\mathrm{h.c.} \,,
\end{equation}
where $v^\mu=(1,\textit{\textbf{0}})$ represents the velocity of the heavy quark.
The Goldstone bosons of the spontaneous breaking of
chiral symmetry can be parametrized as
\begin{align}
u_\mu &= \textrm{i} \left( u^\dagger\partial_\mu u\, -\, u \partial_\mu u^\dagger\right) \,, \qquad
u = \exp \Big( \frac{\textrm{i}\Phi}{\sqrt{2}F} \Big)\,, \nonumber\\
\Phi &=
 \begin{pmatrix}
   {\frac{1}{\sqrt{2}}\pi ^0 +\frac{1}{\sqrt{6}}\eta _8 } & {\pi^+ } & {K^+ }  \\
   {\pi^- } & {-\frac{1}{\sqrt{2}}\pi ^0 +\frac{1}{\sqrt{6}}\eta _8} & {K^0 }  \\
   { K^-} & {\bar{K}^0 } & {-\frac{2}{\sqrt{6}}\eta_8 }  \\
 \end{pmatrix} , \label{eq:u-phi-def}
\end{align}
where $F$ represents the pseudo-Goldstone boson decay constant, and we will
use $F_\pi=92.2\MeV$ for the pions and $F_K=113.0\MeV$ for the
kaons.

The contact term amplitude for the $B_c(2S)(p_a) \to B_c(p_b) \pi(p_c)\pi(p_d)$
process, derived using the chiral effective Lagrangians
in Eq.~\eqref{LagrangianPsiPrimePsipipi} is
\begin{equation}
\label{eq.ChiralAmplitude}
M(s,\cos\theta)
= -\frac{4}{F_\pi^2}( c_1 p_c\cdot p_d +c_2 p_c^0 p_d^0)\,,
\end{equation}
where $s= (p_c+p_d)^2$, and $\theta$ is the
angle between the 3-momentum of the $\pi^+$ in the rest frame of the dipion
system and that of the dipion system in the rest frame of the initial
$B_c(2S)$.
By matching the amplitude in Eq.~\eqref{eq.MultipoleAmplitude} to that in Eq.~\eqref{eq.ChiralAmplitude}, the chiral
effective Lagrangian coupling constants $c_1$ and $c_2$ can be expressed in terms of the chromopolarizability $\alpha_{AB}$ and the parameter $\kappa$,
\begin{align}
c_1&= -\frac{\pi^2\sqrt{m_{B_c(2S)}m_{B_c}}F_\pi^2}{b}\alpha_{B_c(2S)B_c}(4+3\kappa) , \nn\\
c_2&= \frac{12\pi^2\sqrt{m_{B_c(2S)}m_{B_c}}F_\pi^2}{b}\alpha_{B_c(2S)B_c}\kappa\,.\label{eq.Matching}
\end{align}

The strong FSIs between two pseudoscalar mesons can be incorporated
model-independently using dispersion theory. Since the invariant
mass of the pion pair approaches to the $K\bar{K}$ threshold, we
will consider the coupled-channel ($\pi\pi$ and $K\bar
K$) FSI for the dominant $S$-wave component, while for the $D$-wave
we will only take account of the single-channel FSI.

For the $B_c(2S)(p_a) \to B_c(p_b) \pi^+(p_c)\pi^-(p_d)$
processes, the partial-wave expansion of the amplitude including FSI
is expressed as \be \M^\text{full}(s,\cos\theta) = \sum_{l=0}^{\infty}
M_l^\pi(s) P_l(\cos\theta)\, \ee

For the dominant $S$-wave, we consider the two-channel
rescattering effects. The two-channel
unitarity conditions reads
\begin{equation}\label{eq.unitarity2channel}
\textrm{Im}\, \textit{\textbf{M}}_0(s)=2 \textrm{i} T_0^{0\ast}(s)\Sigma(s)
\textit{\textbf{M}}_0(s) \,,
\end{equation}
where the two-dimensional vector $\textit{\textbf{M}}_0(s)$ contains both the $\pi\pi$ and the $K\bar{K}$
final states,
 \begin{equation}
\textit{\textbf{M}}_0(s)=\left( {\begin{array}{*{2}c}
   {M^\pi_0(s)} \\
   {\frac{2}{\sqrt{3}}M^K_0(s)}  \\\end{array}} \right)\,.
 \end{equation}
The two-dimensional matrices $T_0^0(s)$ and $\Sigma(s)$ are defined as
\begin{equation}\label{eq.T00}
T_0^0(s)=
 \left( {\begin{array}{*{2}c}
   \frac{\eta_0^0(s)e^{2\textrm{i}\delta_0^0(s)}-1}{2\textrm{i}\sigma_\pi(s)} & |g_0^0(s)|e^{\textrm{i}\psi_0^0(s)}   \\
  |g_0^0(s)|e^{\textrm{i}\psi_0^0(s)} & \frac{\eta_0^0(s)e^{2\textrm{i}\left(\psi_0^0(s)-\delta_0^0(s)\right)}-1}{2\textrm{i}\sigma_K(s)} \\
\end{array}} \right)
\end{equation}
and
$\Sigma(s)\equiv \text{diag} \big(\sigma_\pi(s)\theta(s-4m_\pi^2),\sigma_K(s)\theta(s-4m_K^2)\big)$.
The $T_0^0(s)$ matrix incorporates three input functions: the
$\pi\pi$ $S$-wave isoscalar phase shift $\delta_0^0(s)$ and the $\pi\pi \to K\bar{K}$ $S$-wave
amplitude $g_0^0(s)=|g_0^0(s)|e^{\textrm{i}\psi_0^0(s)}$ with its modulus and
phase. We will use the parameterization of the $T_0^0(s)$ matrices given in Refs.~\cite{Leutwyler2012,Moussallam2004}. These inputs are used up to $\sqrt{s_0}=1.3\GeV$ (the $f_0(1370)$ resonance is known to have a significant coupling to $4\pi$~\cite{ParticleDataGroup:2024cfk}).
Beyond $s_0$, the phases
$\delta_0^0(s)$ and $\psi_0^0$ are smoothly guided to
2$\pi$ via~\cite{Moussallam2000}
\begin{equation}
\delta(s)=2\pi+(\delta(s_0)-2\pi)\frac{2}{1+(\frac{s}{s_0})^{3/2}}\,.
\end{equation}
The inelasticity $\eta_0^0(s)$ in Eq.~\eqref{eq.T00} is related to
the modulus $|g_0^0(s)|$ by
\begin{equation}
\eta_0^0(s)=\sqrt{1-4\sigma_\pi(s)\sigma_K(s)|g_0^0(s)|^2\theta(s-4m_K^2)}\,.
\end{equation}

The solution to the two-channel
unitarity condition in Eq.~\eqref{eq.unitarity2channel} is given by
\begin{equation}\label{OmnesSolution2channel}
\textit{\textbf{M}}_0(s)=\Omega(s) \textit{\textbf{P}}^{n-1}(s) \,,
\end{equation}
where $\Omega(s)$ satisfies the homogeneous two-channel unitarity
relation
\begin{equation}\label{eq.unitarity2channelhomo}
\textrm{Im}\, \Omega(s)=2\textrm{i} T_0^{0\ast}(s)\Sigma(s) \Omega(s),
\hspace{1cm}  \Omega(0)=\mathbbm{1} \,.
\end{equation}
Numerical results for $\Omega(s)$ are available in
Refs.~\cite{Leutwyler90,Moussallam2000,Hoferichter:2012wf,Daub}.

For the $D$-wave, we consider the single-channel FSI. In the elastic
$\pi\pi$ rescattering region, the partial-wave unitarity conditions
is expressed as
\begin{equation}\label{eq.unitarity1channel}
\textrm{Im}\, M_2(s)= M_2(s)
\sin\delta_2^0(s) e^{-\textrm{i}\delta_2^0(s)}\,.
\end{equation}
In the elastic region, the phase of the $D$-wave isoscalar amplitude $\delta_2^0$ equals the
$\pi\pi$ elastic phase shifts modulo $n\pi$, as required by Watson's
theorem~\cite{Watson1,Watson2}. The Omn\`es function is defined
as~\cite{Omnes}
\begin{equation}\label{Omnesrepresentation}
\Omega_2^0(s)=\exp
\bigg\{\frac{s}{\pi}\int^\infty_{4m_\pi^2}\frac{\textrm{d} x}{x}
\frac{\delta_2^0(x)}{x-s}\bigg\}\,.
\end{equation}
Using the Omn\`es function, the solution of
Eq.~\eqref{eq.unitarity1channel} is given by
\be\label{OmnesSolution1channel}
M_2(s)=\Omega_2^0(s)P_2^{n-1}(s) \,, \ee where
the polynomial $P_2^{n-1}(s)$ is a subtraction function. We use the result of the $D$-wave phase shift $\delta_2^0(s)$ given
in Ref.~\cite{Pelaez}, and extend it smoothly to $\pi$ as $s\to\infty$.

On the other hand, at low energies, the amplitudes $M_2(s)$ and $\textit{\textbf{M}}_0(s)$ should
match the results from chiral perturbation theory. Namely, if one turn off the final-state interactions,
$\Omega_2^0(s)=1$ and $\Omega(0)=\mathbbm{1}$, the subtraction
functions should agree with the partial-wave projections of the chiral result obtained using the Lagrangian in
Eq.~\eqref{LagrangianPsiPrimePsipipi}. For the $S$-wave, this leads to the integral equation:
\begin{equation}\label{M02channel}
\textit{\textbf{M}}_0(s)=\Omega(s) \textit{\textbf{M}}_0^{\chi}(s) \,,
\end{equation}
where
\begin{align}
\textit{\textbf{M}}^{\chi}_0(s)=&\big(
   M_0^{\chi,\pi}(s),
   2/\sqrt{3}\,M_0^{\chi,K}(s)
   \big)^{T}\,, \\
M_0^{\chi,\pi(K)}(s)=&-\frac{2}{F_{\pi(K)}^2}\sqrt{m_{B_c(2S)}m_{B_c}}
\bigg\{c_1 \left(s-2m_{\pi(K)}^2 \right)
+\frac{c_2}{2} \bigg[s+\textit{\textbf{q}}^2\Big(1  -\frac{\sigma_{\pi(K)}^2}{3}
\Big)\bigg]\bigg\}\,.\label{eq.M0chiral}
\end{align}

For the $D$-wave, the integral equation is given by\be\label{M21channel}
M_2^\pi(s)=\Omega_2^0(s) M_2^{\chi,\pi}(s) \,, \ee
where
\begin{align}
M_2^{\chi,\pi}(s)&=\frac{2}{3
F_\pi^2}\sqrt{m_{B_c(2S)}m_{B_c}}\,c_2
|\textit{\textbf{q}}|^2\sigma_\pi^2\,.\label{eq.M2chiral}
\end{align}

The $\pi\pi$ invariant mass distribution for the
$B_c(2S) \rightarrow B_c \pi^+\pi^-$
is expressed as
\begin{equation}
\frac{\textrm{d}\Gamma}{\textrm{d} \sqrt{s}} =
\int_{-1}^1 \frac{\sqrt{s}\,\sigma_\pi |\textit{\textbf{q}}|}{128\pi^3 m_{B_c(2S)}^2}
\left|M_0^\pi+M_2^\pi
P_2(\cos\theta)\right|^2   \textrm{d}\cos\theta\,,\label{eq.pipimassdistribution}
\end{equation}
where the Legendre polynomial $P_2(\cos\theta)=(3\cos^2\theta-1)/2.$

\section{Phenomenological discussion} \label{pheno}

Since for the $B_c(2S) \rightarrow B_c \pi^+\pi^-$ process the CMS Collaboration~\cite{CMS:2020rcj} only provides the normalized dipion invariant mass distributions, we use the theoretical prediction of $\Gamma_{B_c(2S) \rightarrow B_c \pi\pi}=\Gamma_{B_c(2S)} \times  Br(B_c(2S) \rightarrow B_c \pi\pi)=73.1 \times 0.811 ~\text{keV}=59.3 ~\text{keV}$ in Ref.~\cite{Eichten:2019gig} to transform them into the differential widths over the dipion invariant mass, which will be fitted in our study. Note that the theoretical prediction of the decay rate $\Gamma_{B_c(2S) \rightarrow B_c \pi\pi}$ in Ref.~\cite{Eichten:2019gig} is the sum of the decay modes with charged and neutral pions, and we take the decay rate of the neutral pions mode to be half of the decay rate of the charged pions mode.

\begin{figure}
 \begin{center}
  \includegraphics[width=0.8\textwidth]{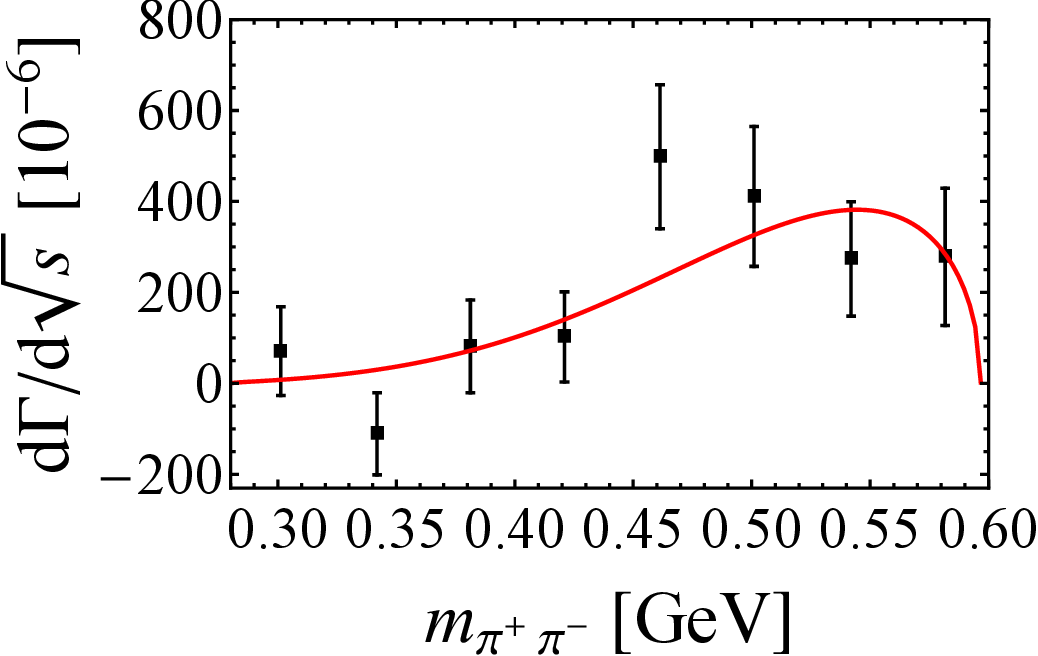}
 \end{center}
 \caption{Fit to the $\pi\pi$ invariant mass distributions in $B_c(2S) \rightarrow B_c \pi^+\pi^-$.
 }
 \label{fig:fitresults}
\end{figure}

The two free parameters are the low-energy coupling constants $c_1$ and $c_2$ in the
chiral Lagrangian~\eqref{LagrangianPsiPrimePsipipi}, which can be expressed
in terms of the chromopolarizability $\alpha_{B_c(2S)B_c}$ and the parameter $\kappa$ as in Eq.~\eqref{eq.Matching}.
By performing the $\chi^2$ fit to the dipion invariant mass distributions of $B_c(2S) \rightarrow B_c \pi^+\pi^-$, we can determine the values of the parameters:
\begin{eqnarray} \label{fit-result}
&& \alpha_{B_c(2S)B_c} =0.66\pm 0.25 \,,\qquad \kappa =-0.02\pm 0.13\,,
\end{eqnarray}
with $\chi^2/{\rm d.o.f }= 6.89/(8-2)=1.15$. The fit results are plotted in Fig.~\ref{fig:fitresults}, and it can be seen that the dipion invariant mass distributions of $B_c(2S) \rightarrow B_c \pi^+\pi^-$ can be well described. In the following, we discuss the fit results in detail.

\begin{figure}
 \begin{center}
  \includegraphics[width=0.8\textwidth]{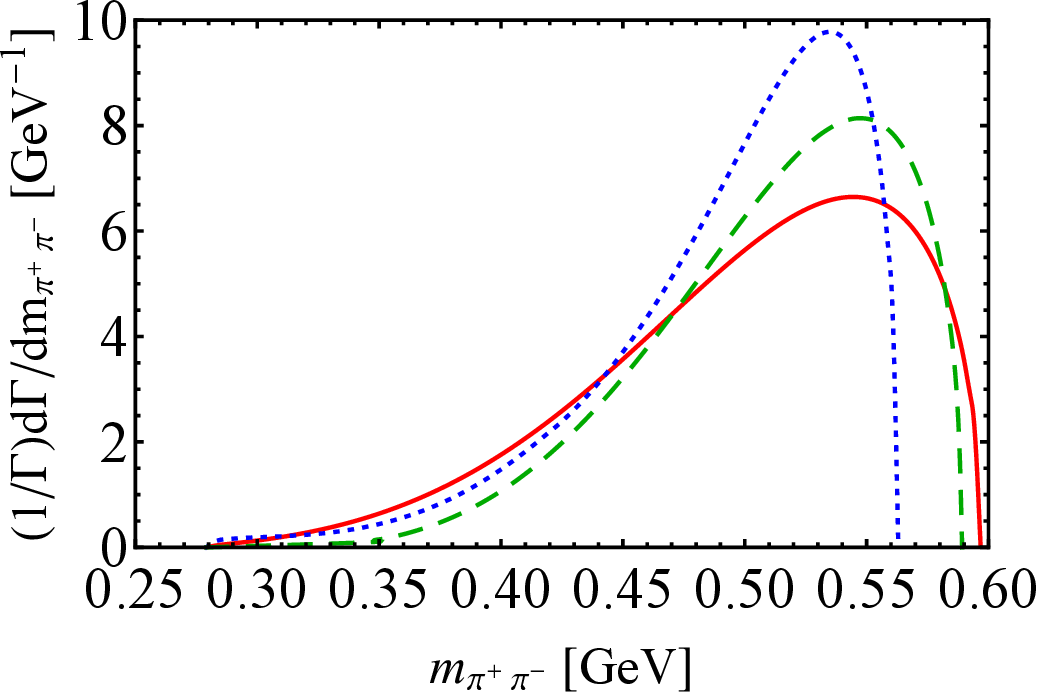}
 \end{center}
 \caption{Normalized dipion mass spectra for the transitions of $\psi(2S) \to J/\psi \pi\pi $ (green dashed)~\cite{Dong:2021lkh}, $B_c(2S) \to B_c \pi\pi $(red sholid), and $\Upsilon(2S) \to \Upsilon \pi\pi $(blue dotted)~\cite{Chen:2019gty}, respectively.
 }
 \label{fig:MpipiNormalized}
\end{figure}

\begin{table}
\caption{\label{tablepar} The transitional chromopolarizabilities $\alpha_{AB}$ and the parameters $\kappa$ extracted from $\psi(2S) \to J/\psi \pi\pi $, $B_c(2S) \to B_c \pi\pi $, and $\Upsilon(2S) \to \Upsilon \pi\pi $, respectively.}
\renewcommand{\arraystretch}{1.2}
\begin{center}
\begin{tabular}{lccc}
\toprule
         & $\psi(2S) \to J/\psi \pi\pi $~\cite{Dong:2021lkh}
         & $B_c(2S) \to B_c \pi\pi $ & $\Upsilon(2S) \to \Upsilon \pi\pi $~\cite{Chen:2019gty}\\
\hline
$|\alpha_{AB}| $ (GeV$^{-3}$)   &   $ 1.81\pm 0.01$   &   $ 0.66\pm 0.25$          & $0.29\pm 0.20 $  \\
$\kappa$   &   $ 0.10\pm 0.01$ & $-0.02\pm 0.13 $  & $1.52\pm 1.17 $ \\
\botrule
\end{tabular}
\end{center}
\renewcommand{\arraystretch}{1.0}
\end{table}

 For the transition chromopolarizability $\alpha_{B_c(2S)B_c}$, first, as shown in Eq.~\eqref{eq.Matching}, it depends on the overall decay rate. In this study, we use the theoretical prediction of the decay rate of $\Gamma_{B_c(2S) \rightarrow B_c \pi\pi}$ provided in Ref.~\cite{Eichten:2019gig} as input, and note that the variations in decay rate predictions across~\cite{Godfrey:2004ya,Eichten:2019gig,Martin-Gonzalez:2022qwd,Gao:2024yvz} are minor. A more precise value of $\alpha_{B_c(2S)B_c}$ can be determined when the experimental data of the decay rate of $\Gamma_{B_c(2S) \rightarrow B_c \pi\pi}$ becomes available in the future. Second, Ref.~\cite{CMS:2020rcj} also provides the normalized dipion invariant mass distributions of the $B_c^\ast(2S) \rightarrow B_c^\ast \pi^+\pi^-$ process, which are compatible with the normalized dipion invariant mass distributions of the $B_c(2S) \rightarrow B_c \pi^+\pi^-$ process within the uncertainties. However, since the $B_c^\ast$ state has not been observed yet and the masses of $B_c^\ast$ and $B_c^\ast(2S)$ have not been measured experimentally~\cite{ParticleDataGroup:2024cfk}, we choose not to fit the data of the normalized dipion mass spectra of the $B_c^\ast(2S) \rightarrow B_c^\ast \pi^+\pi^-$ process.
Nevertheless, since the vector charm-beauty mesons are the spin partners of the pseudoscalar ones, their polarizabilities should be the same as those of the pseudoscalar mesons at leading order, as a consequence of heavy quark spin symmetry. Also note that, theoretically, the decay rates of $\Gamma_{B_c(2S) \rightarrow B_c \pi^+\pi^-}$ and $\Gamma_{B_c^\ast(2S) \rightarrow B_c^\ast \pi^+\pi^-}$ are expected to be similar, and both the mass differences between $B_c$ and $B_c^\ast$ and the mass difference between $B_c(2S)$ and $B_c^\ast(2S)$ are expected to be below a few percent~\cite{Godfrey:2004ya,Eichten:2019gig,Martin-Gonzalez:2022qwd,Gao:2024yvz}.
 Therefore, it is expected that the values of chromopolarizabilities $\alpha_{B_c(2S)B_c}$ and $\alpha_{B_c^\ast(2S)B_c^\ast}$ are of the same order of magnitude. Third, since both in mass and in size, the $c\bar{b}$ states are intermediate between the $c\bar{c}$ and $b\bar{b}$ states, it is natural to expect that the diagonal and nondiagonal (transitional) chromopolarizabilities of the $c\bar{b}$ states should lies between corresponding diagonal and nondiagonal chromopolarizabilities of the $c\bar{c}$ and $b\bar{b}$ states, respectively. In Table~\ref{tablepar}, the transitional chromopolarizabilities $\alpha_{AB}$ and the parameters $\kappa$ extracted from $\psi(2S) \to J/\psi \pi\pi $~\cite{Dong:2021lkh}, $B_c(2S) \to B_c \pi\pi $, and $\Upsilon(2S) \to \Upsilon \pi\pi $~\cite{Chen:2019gty}, respectively, are given. Assuming $\alpha_{B_c(2S)B_c} \simeq \alpha_{B_c^\ast(2S)B_c^\ast}$, we observe the following hierarchy: $|\alpha_{\Upsilon(2S)\Upsilon(1S)}| < |\alpha_{B_c^{(\ast)}(2S)B_c^{(\ast)}}| < |\alpha_{\psi(2S)\psi}|$, which agrees with above expectation for the transition $2^3S_1 \rightarrow 1^3S_1 \pi\pi$. Note that in Ref.~\cite{Ferretti:2018ojb}, the bottomonium diagonal chromopolarizabilities are calculated by considering them as pure Coulombic systems. Based on the result that the value of $|\alpha_{\Upsilon(2S)\Upsilon(2S)}|$ lies in the range of $(23\sim 33)$ GeV$^{-3}$, several possible $\Upsilon(2S)$-$N$ bound states are predicted~\cite{Ferretti:2018ojb}. It is natural to expect that $|\alpha_{B_c^\ast(2S)B_c^\ast(2S)}| > |\alpha_{\Upsilon(2S)\Upsilon(2S)}|$, and therefore the $B_c^\ast(2S)$-$N$ bound states may also exit. Also note that in Ref.~\cite{Dong:2021lkh}, assuming that the external current couples to a single charm (anti-) quark only, the authors estimate the ratio
$ \xi \equiv \alpha_{J/\psi J/\psi}/ \alpha_{\psi(2S)J/\psi} \sim \mathcal O(10)$ through study of the overlap integral of the spatial wave functions of the initial- and final-state charmonia. Furthermore, they find that even with a conservative estimate of $ \xi =2$, it is possible for two $J/\psi$ mesons to form a bound state.
If a significant contribution to the chromopolarizabilities of the $B_c$ states arises from the coupling of the charm (anti-) quark, one might also estimate that the ratio $\alpha_{B_c^\ast B_c^\ast}/ \alpha_{B_c^{\ast}(2S)B_c^{\ast}} \sim \mathcal O(10)$. As shown in Table~\ref{tablepar}, the value of $|\alpha_{B_c^{(\ast)}(2S)B_c^{(\ast)}}|$ is about one third of the value of $\alpha_{\psi(2S)J/\psi}$, suggesting that the value of $\alpha_{B_c^\ast B_c^\ast}$ may also be large enough to form a two $B_c^\ast$ bound state.

From Eqs.~\eqref{eq.Matching},~\eqref{eq.M0chiral}, and~\eqref{eq.M2chiral}, it is clear that the parameter $\kappa$ characterizes the $\pi\pi$ $D$-wave contribution. As shown in Table~\ref{tablepar}, the central values of the results indicate the hierarchy $|\kappa_{B_c(2S)B_c}| < |\kappa_{\psi(2S)\psi}| < |\kappa_{\Upsilon(2S)\Upsilon(1S)}| $. In Fig.~\ref{fig:fitresults}, we plot the normalized dipion mass spectra for the transitions of $\psi(2S) \to J/\psi \pi\pi $ (green dashed), $B_c(2S) \to B_c \pi\pi $(red sholid), and $\Upsilon(2S) \to \Upsilon \pi\pi $(blue dotted), respectively, based on the unified theoretical framework employing the chiral effective theory and the dispersive theory. It is observed that, compared to the $\psi(2S) \to J/\psi \pi\pi $ and the $\Upsilon(2S) \to \Upsilon \pi\pi $ cases, the line shape of the normalized dipion mass spectra of $B_c(2S) \to B_c \pi\pi $ is more low-fat. We have checked that the $\pi\pi$ $D$-wave contribution to the total rate is negligible for the $B_c(2S) \to B_c \pi\pi $ process and accounts for less than two percent of the total rate in the $\psi(2S) \to J/\psi \pi\pi $ process.

\section{Conclusions}
\label{conclu}

We employed dispersion theory to investigate the $\pi\pi$ FSI in the
$B_c(2S) \to B_c \pi^+\pi^-$ transition. By fitting the $\pi\pi$ mass spectra data, the values of the transitional chromopolarizability $\alpha_{B_c(2S)B_c}$
and the parameter $\kappa$ are determined. Our results indicate that the transitional chromopolarizability of the $c\bar{b}$ state lies between the transitional chromopolarizabilities of the $c\bar{c}$ and $b\bar{b}$ states.
Additionally, the parameter $\kappa$, which accounts for the $\pi\pi$ $D$-wave contribution, is tiny. The findings of this study provide valuable insights into the chromopolarizability of $B_c$ states and have potential applications in the investigation of multiquark hadrons containing $c\bar{b}$.

\section*{Acknowledgments}

This work is supported in part by the Fundamental Research Funds
for the Central Universities under Grants No.~FRF-BR-19-001A.


\begin{thebibliography}{99}


\bibitem{CDF:1998ihx}
F.~Abe \textit{et al.} [CDF],
Observation of the $B_c$ meson in $p\bar{p}$ collisions at $\sqrt{s} = 1.8$ TeV,
Phys. Rev. Lett. \textbf{81} (1998), 2432-2437.

\bibitem{CDF:1998axz}
F.~Abe \textit{et al.} [CDF],
Observation of $B_c$ mesons in $p\bar{p}$ collisions at $\sqrt{s} = 1.8$ TeV,
Phys. Rev. D \textbf{58} (1998), 112004.


\bibitem{ATLAS:2014lga}
G.~Aad \textit{et al.} [ATLAS],
Observation of an Excited $B_c^\pm$ Meson State with the ATLAS Detector,
Phys. Rev. Lett. \textbf{113} (2014) no.21, 212004.


\bibitem{CMS:2019uhm}
A.~M.~Sirunyan \textit{et al.} [CMS],
Observation of Two Excited B$^+_\mathrm{c}$ States and Measurement of the B$^+_\mathrm{c}$(2S) Mass in pp Collisions at $\sqrt{s} =$ 13 TeV,
Phys. Rev. Lett. \textbf{122} (2019) no.13, 132001.

\bibitem{LHCb:2019bem}
R.~Aaij \textit{et al.} [LHCb],
Observation of an excited $B_c^+$ state,
Phys. Rev. Lett. \textbf{122} (2019) no.23, 232001.


\bibitem{CMS:2020rcj}
A.~M.~Sirunyan \textit{et al.} [CMS],
Measurement of B$_\mathrm{c}$(2S)$^+$ and B$_\mathrm{c}^*$(2S)$^+$ cross section ratios in proton-proton collisions at $\sqrt{s} =$ 13 TeV,
Phys. Rev. D \textbf{102} (2020) no.9, 092007.



\bibitem{Voloshin:1980zf}
  M.~B.~Voloshin and V.~I.~Zakharov,
  Measuring QCD Anomalies in Hadronic Transitions Between Onium States,
  Phys.\ Rev.\ Lett.\  {\bf 45}, 688 (1980).

\bibitem{Novikov:1980fa}
  V.~A.~Novikov and M.~A.~Shifman,
  Comment on the $\psi^\prime \to J/\psi \pi \pi$ Decay,
  Z.\ Phys.\ C {\bf 8}, 43 (1981).

\bibitem{Kuang:1981se}
Y.~P.~Kuang and T.~M.~Yan,
Predictions for Hadronic Transitions in the B anti-B System,
Phys. Rev. D \textbf{24}, 2874 (1981).

\bibitem{Kuang:2006me}
Y.~P.~Kuang,
QCD multipole expansion and hadronic transitions in heavy quarkonium systems,
Front. Phys. China \textbf{1}, 19-37 (2006).




\bibitem{Eichten:1994gt}
E.~J.~Eichten and C.~Quigg,
Mesons with beauty and charm: Spectroscopy,
Phys. Rev. D \textbf{49}, 5845-5856 (1994).

\bibitem{Godfrey:2004ya}
S.~Godfrey,
Spectroscopy of $B_c$ mesons in the relativized quark model,
Phys. Rev. D \textbf{70}, 054017 (2004).


\bibitem{Eichten:2019gig}
E.~J.~Eichten and C.~Quigg,
Mesons with Beauty and Charm: New Horizons in Spectroscopy,
Phys. Rev. D \textbf{99}, no.5, 054025 (2019).


\bibitem{Martin-Gonzalez:2022qwd}
B.~Mart\'\i{}n-Gonz\'alez, P.~G.~Ortega, D.~R.~Entem, F.~Fern\'andez and J.~Segovia,
Toward the discovery of novel Bc states: Radiative and hadronic transitions,
Phys. Rev. D \textbf{106}, no.5, 054009 (2022).


\bibitem{Li:2023wgq}
X.~J.~Li, Y.~S.~Li, F.~L.~Wang and X.~Liu,
Spectroscopic survey of higher-lying states of $B_c$ meson family,
Eur. Phys. J. C \textbf{83}, no.11, 1080 (2023).

\bibitem{Gao:2024yvz}
Z.~b.~Gao, Y.~y.~Fan, H.~Chen and C.~q.~Pang,
M1 radiative and spin-nonflip \ensuremath{\pi}\ensuremath{\pi} transitions of Bc states in the Cornell potential model,
Phys. Rev. D \textbf{110}, no.3, 034003 (2024).



\bibitem{Chen:2015jgl}
Y.~H.~Chen, J.~T.~Daub, F.~K.~Guo, B.~Kubis, U.~G.~Mei\ss{}ner and B.~S.~Zou,
Effect of $Z_b$ states on $\Upsilon(3S)\to\Upsilon(1S)\pi\pi$ decays,
Phys. Rev. D \textbf{93}, no.3, 034030 (2016).


\bibitem{Chen:2016mjn}
Y.~H.~Chen, M.~Cleven, J.~T.~Daub, F.~K.~Guo, C.~Hanhart, B.~Kubis, U.~G.~Mei\ss{}ner and B.~S.~Zou,
Effects of $Z_b$ states and bottom meson loops on $\Upsilon(4S) \to \Upsilon(1S,2S) \pi^+\pi^-$ transitions,
Phys. Rev. D \textbf{95}, no.3, 034022 (2017).


\bibitem{Chen:2019gty}
Y.~H.~Chen and F.~K.~Guo,
Chromopolarizabilities of bottomonia from the $\Upsilon(2S,3S,4S) \to \Upsilon(1S,2S)\pi\pi$ transitions,
Phys. Rev. D \textbf{100}, no.5, 054035 (2019).


\bibitem{Chen:2019hmz}
Y.~H.~Chen,
Chromopolarizability of Charmonium and $\ensuremath{\pi}\ensuremath{\pi}$ Final State Interaction Revisited,
Adv. High Energy Phys. \textbf{2019}, 7650678 (2019).







\bibitem{Sibirtsev:2005ex}
  A.~Sibirtsev and M.~B.~Voloshin,
The Interaction of slow $J/\psi$ and $\psi^\prime$ with nucleons,
  Phys.\ Rev.\ D {\bf 71}, 076005 (2005).

\bibitem{Voloshin:2007dx}
  M.~B.~Voloshin,
  Charmonium,'
  Prog.\ Part.\ Nucl.\ Phys.\  {\bf 61}, 455 (2008).

\bibitem{Dubynskiy:2008mq}
  S.~Dubynskiy and M.~B.~Voloshin,
  Hadro-Charmonium,
  Phys.\ Lett.\ B {\bf 666} (2008) 344.


\bibitem{Eides:2015dtr}
  M.~I.~Eides, V.~Y.~Petrov and M.~V.~Polyakov,
  Narrow nucleon-$\psi(2S)$ bound state and LHCb pentaquarks,
  Phys.\ Rev.\ D {\bf 93}, 054039 (2016);
  M.~I.~Eides, V.~Y.~Petrov and M.~V.~Polyakov,
  Pentaquarks with hidden charm as hadroquarkonia,
  Eur.\ Phys.\ J.\ C {\bf 78}, 36 (2018).


\bibitem{Tsushima:2011kh}
  K.~Tsushima, D.~H.~Lu, G.~Krein and A.~W.~Thomas,
  $J/\Psi$-nuclear bound states,
  Phys.\ Rev.\ C {\bf 83}, 065208 (2011).


\bibitem{Dong:2021lkh}
X.~K.~Dong, V.~Baru, F.~K.~Guo, C.~Hanhart, A.~Nefediev and B.~S.~Zou,
Is the existence of a $J/\ensuremath{\psi}J/\ensuremath{\psi}$ bound state plausible?,
Sci. Bull. \textbf{66}, no.24, 2462-2470 (2021)
doi:10.1016/j.scib.2021.09.009
[arXiv:2107.03946 [hep-ph]].


\bibitem{Ferretti:2018kzy}
  J.~Ferretti,
  $\eta_{\rm c}$- and $J/\psi$-isoscalar meson bound states in the hadro-charmonium picture,
  Phys.\ Lett.\ B {\bf 782}, 702 (2018)

\bibitem{Ferretti:2018ojb}
  J.~Ferretti, E.~Santopinto, M.~Naeem Anwar and M.~A.~Bedolla,
  The baryo-quarkonium picture for hidden-charm and bottom pentaquarks and LHCb $P_{\rm c}(4380)$ and $P_{\rm c}(4450)$ states,
  Phys.\ Lett.\ B {\bf 789}, 562 (2019).




\bibitem{Dong:2022rwr}
X.~K.~Dong, F.~K.~Guo, A.~Nefediev and J.~T.~Castell\`a,
Chromopolarizabilities of fully heavy baryons,
Phys. Rev. D \textbf{107}, no.3, 034020 (2023)
doi:10.1103/PhysRevD.107.034020
[arXiv:2211.14100 [hep-ph]].


\bibitem{Voloshin:2004un}
  M.~B.~Voloshin,
  Quarkonium chromo-polarizability from the decays
  $ J/\psi(\Upsilon) \to \pi \pi l^+ l^-$,
  Mod.\ Phys.\ Lett.\ A {\bf 19}, 665 (2004).

 \bibitem{Brambilla:2015rqa}
  N.~Brambilla, G.~Krein, J.~Tarr\'us Castell\`a and A.~Vairo,
  Long-range properties of $1S$ bottomonium states,
  Phys.\ Rev.\ D {\bf 93}, 054002 (2016).


\bibitem{Mannel}
  T.~Mannel and R.~Urech,
  Hadronic decays of excited heavy quarkonia,
  Z.\ Phys.\ C {\bf 73}, 541 (1997).




\bibitem{Leutwyler2012}
  I.~Caprini, G.~Colangelo, and H.~Leutwyler,
  Regge analysis of the $\pi\pi$ scattering amplitude,
  Eur.\ Phys.\ J.\ C {\bf 72}, 1860 (2012).


\bibitem{Moussallam2004}
  P.~B\"uttiker, S.~Descotes-Genon, and B.~Moussallam,
  A new analysis of pi K scattering from Roy and Steiner type equations,
  Eur.\ Phys.\ J.\ C {\bf 33}, 409 (2004).



\bibitem{ParticleDataGroup:2024cfk}
S.~Navas \textit{et al.} [Particle Data Group],
Review of particle physics,
Phys. Rev. D \textbf{110} (2024) no.3, 030001.




\bibitem{Moussallam2000}
  B.~Moussallam,
  N(f) dependence of the quark condensate from a chiral sum rule,
  Eur.\ Phys.\ J.\ C {\bf 14}, 111 (2000).

\bibitem{Leutwyler90}
  J.~F.~Donoghue, J.~Gasser, and H.~Leutwyler,
  The Decay of a Light Higgs Boson,
  Nucl.\ Phys.\ B {\bf 343}, 341 (1990).

\bibitem{Daub}
  J.~T.~Daub, C.~Hanhart, and B.~Kubis,
  A model-independent analysis of final-state interactions in $\bar B_{d/s}^0 \to J/\psi \pi \pi$,
  J.~High Energy Phys.\ {\bf 02} (2016) 009.

\bibitem{Hoferichter:2012wf}
  M.~Hoferichter, C.~Ditsche, B.~Kubis, and U.-G.~Mei{\ss}ner,
  Dispersive analysis of the scalar form factor of the nucleon,
  J.~High Energy Phys.\ {\bf 06} (2012) 063.



\bibitem{Watson1}
  K.~M.~Watson,
  The Effect of Final State Interactions on Reaction Cross Sections,
  Phys.\ Rev.\  {\bf 88}, 1163 (1952).

\bibitem{Watson2}
  K.~M.~Watson,
  Some general relations between the photoproduction and scattering of pi
  mesons,
  Phys.\ Rev.\  {\bf 95}, 228 (1954).

\bibitem{Omnes}
  R.~Omn\`es,
  On the Solution of certain singular integral equations of quantum field theory,
  Nuovo Cim.\  {\bf 8}, 316 (1958).



\bibitem{Pelaez}
  R.~Garc\'ia-Mart\'in, R.~Kami\'nski, J.~R.~Pel\'aez, J.~Ruiz de Elvira, and F.~J.~Yndur\'ain,
  The pion-pion scattering amplitude. IV: Improved analysis with once subtracted Roy-like equations up to 1100 MeV,
  Phys.\ Rev.\  D {\bf 83}, 074004 (2011).







\end{thebibliography}
\end{document}